\documentclass[prb,epsfig,twocolumn,showpacs,amsmath,amssymb,superscriptaddress]{revtex4}
\usepackage{epsfig}  
\usepackage{dcolumn}  
\usepackage{bm}
\usepackage{braket} 
\usepackage[latin1]{inputenc}                                 
\usepackage{color}                                            
\usepackage{calc}                                             
\usepackage{hhline}                                           

    \def\FeIII{Fe$^{3+}$ }   
 
 \def\muB{\mu_{\text{B}}}
\begin{document}

\title{Determination of the single-ion anisotropy energy in a $S$ =
5/2 kagome antiferromagnet using x-ray absorption spectroscopy.}
\author{M.~A.~de~Vries} \email{m.a.devries@physics.org}
\affiliation{Laboratory for Quantum Magnetism, \'Ecole Polytechnique
F\'ed\'erale de Lausanne, Station 3, CH-1015 Lausanne, Switzerland}
\affiliation{ CSEC and School of Chemistry, The University of
Edinburgh, Edinburgh EH9 3JZ, UK} \author{T.~K.~Johal} \affiliation{
Daresbury Laboratory, Warrington WA4 4AD, UK} \author{A.~Mirone}
\affiliation{ESRF, 6 rue Jules Horowitz, F-38000 Grenoble, France}
\author{J.~S.~Claydon} \affiliation{Department of Engineering
Materials, University of Sheffield, Sheffield S1 3JD, UK}
\author{G.~J.~Nilsen} \affiliation{Laboratory for Quantum Magnetism,
\'Ecole Polytechnique F\'ed\'erale de Lausanne, Station 3, CH-1015
Lausanne, Switzerland}   \affiliation{ CSEC and School of Chemistry,
The  University of Edinburgh, Edinburgh EH9 3JZ, UK}
\author{H.~M.~R\o{}nnow}\affiliation{Laboratory for Quantum Magnetism,
\'Ecole Polytechnique F\'ed\'erale de Lausanne, Station 3, CH-1015
Lausanne, Switzerland}   \author{G.~van der Laan}    \affiliation{
Daresbury Laboratory, Warrington WA4 4AD, UK} \affiliation{  Diamond
Light Source, Chilton, Didcot OX11 0DE, UK} \author{A.~Harrison}
\affiliation{ CSEC and School of Chemistry, The  University of
Edinburgh, Edinburgh EH9 3JZ, UK}  \affiliation{Institut
Laue-Langevin, 6 rue Jules Horowitz, F-38000 Grenoble, France}
\date{\today}

\begin{abstract} We report x-ray absorption and x-ray linear dichroism
measurements at the Fe $L_{2,3}$ edges of the geometrically frustrated
systems of potassium and hydronium iron jarosite. Comparison with
simulated spectra, involving ligand-field multiplet calculations
modelling the $3d$-$2p$ hybridization between the iron ion and the
oxygen ligands, has yielded accurate estimates for the ligand
metal-ion hybridization and the resulting single-ion crystal field
anisotropy energy. Using this method we provide an experimentally
verified scenario for the appearance of a single-ion anisotropy in
this nominally high-spin $3d^5$ orbital singlet $^6S$ system, which
accounts for features of the spin-wave dispersion in the long-range
ordered ground state of potassium iron jarosite. 
\end{abstract}


\maketitle{}
\section{Introduction}  
\begin{figure}[htbp] \epsfig{file = 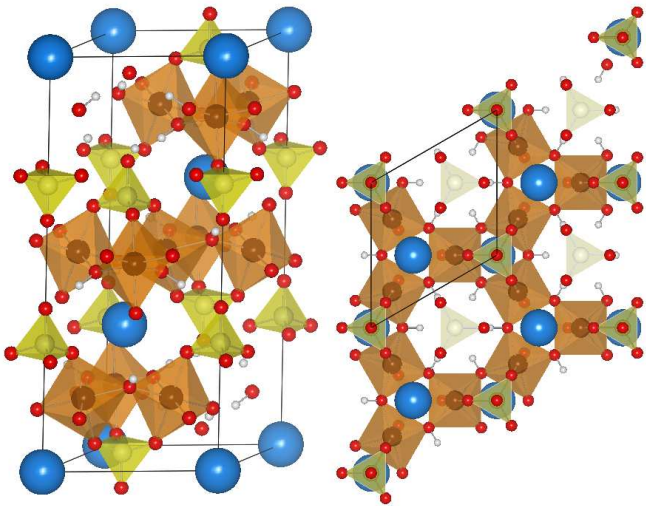, width=3.4in}
\caption{(color online) The jarosite structure viewed perpendicular to
the $c$ axis (left picture) and along the $c$ axis (right picture),
revealing the kagome network of magnetic ions. The $M$O$_6$ octahedra
are drawn in transparent brown with the $M^{3+}$ ions at the center in
dark brown. The SO$_4$ tetrahedra are transparent yellow. The large
spheres at the origin of the unit cells in blue are the $A$ site ions,
and the hydrogen of the OH groups are drawn as small white spheres.  }
\label{figure:Kjar}
\end{figure}

The jarosite group of minerals has received
much attention for being close to ideal realizations of the kagome
antiferromagnet. The kagome topology, which is named after a Japanese
basket weaving pattern, frustrates the antiferromagnetic N\'eel
ordering~\cite{Diep:04}. Due to the resulting ``underconstraint''
neither in the quantum case of $S$ = 1/2 nor in the classical limit is
the kagome antiferromagnet expected to show a symmetry-breaking
transition, even at  $T=0$ K \cite{Reimers:91, Chalker:92,
Reimers:93}. Physical realizations of this system are highly valued
because they allow for the investigation of the Mott insulating phase
in the absence of N\'eel-like magnetic order.  The jarosite group of
general stoichiometry $AM_3$(SO$_4$)$_2$(OH)$_6$, where $A$ can
be---amongst others---K, Na, Rb, Ag, NH$_3$, or H$_3$O, provides
physical models for the kagome antiferromagnet with spin $S$ = 3/2 and
5/2, where the $M$ site is occupied by trivalent Cr~\cite{KerenJar:96,
Inami:01, Morimoto:03} and Fe~\cite{Takano:68, Townsend:86, Wills:96,
WillsThesis, KerenJar:96, WillsKjar:00, WillsKjar:01, Grohol:05},
respectively. $M = $ V is a $S$ = 1 system with ferromagnetic
near-neighbor interactions~\cite{Grohol:02, Grohol:03van}.
Fig.~\ref{figure:Kjar} illustrates the jarosite structure of well
separated kagome layers consisting of $M$O$_6$ octahedra. Potassium
iron jarosite is representative for most of the iron analogues. In
this system a transition to a non-collinear long-range ordered state
is observed at 64~K, despite that the magnitude of the Curie-Weiss
temperature, $\Theta_{\rm{CW}}$, is as high as
$-$800~K~\cite{Grohol:03, WillsKjar:00}. The ground state has shown to
be a so-called $q=0$ state of positive chirality, in which the spins
lie within the kagome plane and all point either in or out of each
shared kagome triangle~\cite{Townsend:86, Inami:00, WillsKjar:01,
Inami:03, Grohol:05}. The large figure $\Theta_{\text{CW}}/T_{\rm{N}}
= 12.5$ is an indication of relatively strong geometric
frustration~\cite{Ramirez:94}, but clearly the observed long-range
ordered ground state points to additional terms in the Hamiltonian of
this kagome antiferromagnet, beyond near-neighbor exchange. It has
been shown that the $q=0$ ground state can arise due to magnetic
anisotropies such as easy-plane single-ion crystal field (CF)
anisotropy $D_z \hat{S}_z^2 - E_{xy} (\hat{S}_{x}^2 - \hat{S}_y^2),
$~\cite{Nishiyama:03, Yildirim:06} where $D_z$ is the zero-field
splitting parameter, or a Dzyaloshinsky-Moriya interaction
(DMI)~\cite{Elhajal:02,Yildirim:06} $D_{ij} \hat{S}_i \times
\hat{S}_j$. The DMI term has shown to be symmetry allowed on the
kagome lattice and has been argued to prevail over the CF anisotropy
in the spherical ($L=0$) 3$d^5$ \FeIII ion~\cite{Elhajal:02}.  Fits to
the spin-wave dispersion curves measured with neutron spectroscopy on
single crystal~\cite{Matan:06} and powder samples~\cite{Coomer:06} of
potassium jarosite have found to be slightly better in a model with
DMI compared to the model with CF anisotropy. The evidence so far has,
however, been inconclusive. 

Given the similarity in crystal structure of potassium (iron) jarosite
and hydronium (iron) jarosite the above effects could expected to be
equally important in the latter. However, hydronium jarosite has a
spin-glass ground state with a freezing temperature of $T_{g} =
17(2)$~K~\cite{Bisson:unpub07}. The Weiss temperature for hydronium
jarosite is less accurately known ($-$700 to $-$1400~K) because the
inverse susceptibility does not enter a linear regime sufficiently far
below the decomposition temperature of the
compound~\cite{Wills:96}. It has been suggested that the spin-glass
state is the result of proton transfer from the hydronium groups to
the Fe-(O$_{\text{eq}}$H)-Fe super-exchange mediating hydroxy
groups~\cite{Grohol:03}. One indication that this could happen might
be that the OH groups at the crystallographically analogous location
in the related kagome compound zinc paratacamite can form a muonium
state by chemically binding a positive muon~\cite{Mendels:07}.  The
aim of the work described here is to investigate the origins of the
magnetic anisotropy (of either single-ion or DMI character) in the
iron jarosites, and to understand the essential differences between
the salts such as potassium jarosite with a long range ordered ground
state and hydronium jarosite.  

Using M\"{o}ssbauer spectroscopy~\cite{Afanasev:74, Bonville:06} the
spin state of the Fe$^{3+}$ in the iron jarosites was found to be $S$
= 5/2, thus nominally high-spin $3d^5$. The free-ion 3$d^5$ state is
$^6S$, i.e. $L = 0$. Hence, there is no orbital angular momentum due
to the spin-orbit coupling in the presence of a crystal field as long
as the system is high-spin 3$d^5$. This holds at least up to third
order perturbation theory of the trigonal crystal-field distortions,
spin-spin and spin-orbit couplings\footnote{It has been
shown~\cite{Watanabe:57} that in 4th, 5th and 6th order perturbation
theory of the crystal field and spin-spin interactions a small orbital
angular momentum and single-ion anisotropy arises.}.  Even in a
$D_{4h}$ crystal field the 3$d^5$ $S$ = 5/2 spin should therefore be
practically isotropic apart from the possible effects of dipole-dipole
interactions. Both DMI and single-ion anisotropy arise only when the
spin-orbit operator $L \cdot S$ can induce a finite orbital angular
momentum~\cite{Moriya:60,Watanabe:57}. That $\langle L \rangle > 0$ in
the iron jarosites is clear from Curie-Weiss fits to the high
temperature magnetic susceptibility which indicate $\mu_{\text{eff}}$
between 6.2 and 6.7 $\muB$ in potassium iron jarosite~\cite{Grohol:05}
and 6.6(2) $\muB$ in the hydronium
analogue~\cite{Wills:96,WillsH3Ojar:00} (compared to a spin only value
of the effective moment of 5.92 $\muB$). Clearly, this opens up the
possibility for both DMI and single ion anisotropies. However, the
origin of the orbital angular momentum itself has so far received
little attention. 

A very strong crystal field, such as in iron
phthalocyanine~\cite{Thole:88Spinmix} can give rise to a low-spin
ground state in which the spin-orbit coupling will in general
reinstate part of the orbital angular momentum of the free-ion
configuration which is, in zero order, quenched by the crystal
field. It could be that in weaker crystal fields, such as those
present in iron oxides, some orbital angular momentum appears due to a
mixing in of the low-spin state into the ground state. The crystal
field alone does, however, not provide matrix elements between
high-spin and low-spin states. The most likely explanation is
therefore that an Fe$^{2+}$ configuration mixes in due to charge
transfer~\cite{Zaanen:85} from the surrounding ligands. For hematite,
e.g., the effective valence of the Fe ion has been estimated at 2.43
due to charge transfer from the oxygen ligands~\cite{Coey:71}. These
effects have earlier been shown to explain zero field splitting in
Mn$^{2+}$ and Fe$^{3+}$ substituted
crystals~\cite{Wan-Lun:89,Wan-Lun:94} and in the Mn$^{2+}$ compounds
MnF$_3$~\cite{Fransisco:88}, MnPS$_3$ and
MnPSe$_3$~\cite{Jeevanandam:99}, despite the lower covalence of the
fluoride and phosphorus-containing ligands compared to oxygen. 

Transition metal $2p \to 3d$ x-ray spectroscopy studies have played a
key role in the understanding of the effect of charge-transfer and
crystal-field in determining the electronic and magnetic properties in
transition metals and their compounds~\cite{Zaanen:85, vanderLaan:86,
FdG:94Rev, FdG:05Rev}. We have measured the x-ray absorption and x-ray
linear dichroism  (XLD) at the Fe $L_{2,3}$ ($2p \to 3d$) edges on
single-crystalline and powder samples of potassium and hydronium iron
jarosite. By comparison with atomic multiplet calculations, taking
into account ligand-field effects and multiple ionic configurations,
we obtain accurate values for the charge transfer and single-ion
anisotropy in the $S$ = 5/2 kagome antiferromagnet. The results are
compared to magnetic susceptibility and neutron spectroscopy
measurements on in particular potassium iron jarosite.

\section{Experimental} Single crystals of potassium iron jarosite,
KFe$_{3}$(OH)$_{6}$(SO$_{4}$)$_{2}$, were grown using a hydrothermal
reduction-oxidation method as described in
Ref.~\onlinecite{Grohol:03}; 4.88~g (28.0~mmol) K$_2$SO$_4$ and 2.2~mL
(40~mmol) H$_2$SO$_4$ were dissolved in 50~mL distilled water, and
transferred into a 125~mL PTFE liner of a stainless steel bomb. 0.56~g
(10~mmol) iron wire with a diameter of 2~mm was added to the solution.
The bomb was placed in an oven at 202$^{\circ}$C for 4~days, then
cooled down to room temperature at a rate of 0.3$^{\circ}$C/min. The
precipitate was washed, filtered and dried, yielding 0.37~g, which is
22\% based on Fe. In order to obtain sufficiently large crystals it
was important that the surface area of the solution exposed to air
inside the vessel was minimized. This was achieved using a glass
container which fitted inside the PTFE liner, with only a small hole
at the top. Using this method single crystals of up to 0.7~mm in
diameter were obtained, which was sufficiently large for the
application in polarized x-ray spectroscopy. A number of crystals were
characterized using single-crystal x-ray diffraction. These crystals
were identified as potassium iron jarosite and face-indexed.

The original solvothermal synthesis
method~\cite{Dutrizac:76,WillsThesis} for the jarosites was used for
the preparation of H$_3$OFe$_{3}$(OH)$_{6}$(SO$_{4}$)$_{2}$. In this
preparation 6.6~g (22~mmol) of Fe$_2$(SO$_4$)$_3\cdot$5H$_2$O was
dissolved in 50~mL water. This solution was transferred to a 125~mL
PTFE liner of a stainless steel bomb. The solution was heated in the
bomb to 140$^{\circ}$C for 12 hours. The hydronium iron jarosite which
precipitated during the reaction was washed, filtered and dried,
yielding $\sim$0.27~g of product. 

Polarized soft x-ray absorption measurements were carried out using
the liquid helium cryostat, housed in the high-field superconducting
magnet on ID08 at the European Synchrotron Radiation Facility (ESRF)
in Grenoble, France. Total-electron-yield spectra were measured by
recording the drain current from the sample as a function of  photon
energy in the region of the Fe $L_{2,3}$ absorption edges. Potassium
iron jarosite is strongly insulating,  so the resonance enhanced
total-electron-yield signal at the iron edge is much smaller than the
signal from the surrounding metal of the sample holder. To avoid this
source of background intensity a potassium iron jarosite single
crystal of 0.7~mm diameter was attached using silver glue to the end
of a graphite tip, which in turn was glued to the end of a standard
ESRF sample holder.  The sample was aligned in the x-ray beam with the
crystallographic $c$ axis pointing in the horizontal plane and
perpendicular to the incident beam. This allowed us to measure
absorption spectra with the incident x-ray  polarization both
perpendicular and parallel to the $c$ axis. Spectra were recorded with
alternating horizontal and vertical  polarization, and for each
polarization averages were taken over at least eight scans. In this
way the polarization dependent spectra were taken at a number of
temperatures between 20 and 290~K, in zero external field.

Isotropic absorption spectra at the Fe $L_{2,3}$ edges were obtained
on beamline 5U.1 of the Synchrotron Radiation Source (SRS) at
Daresbury Laboratory, UK, using potassium and hydronium iron jarosite
powder dispersed over double-sided UHV-compatible carbon tape.

\section{Results} 
\begin{figure}[htbp] \vspace{0.3cm}
\begin{center} \epsfig{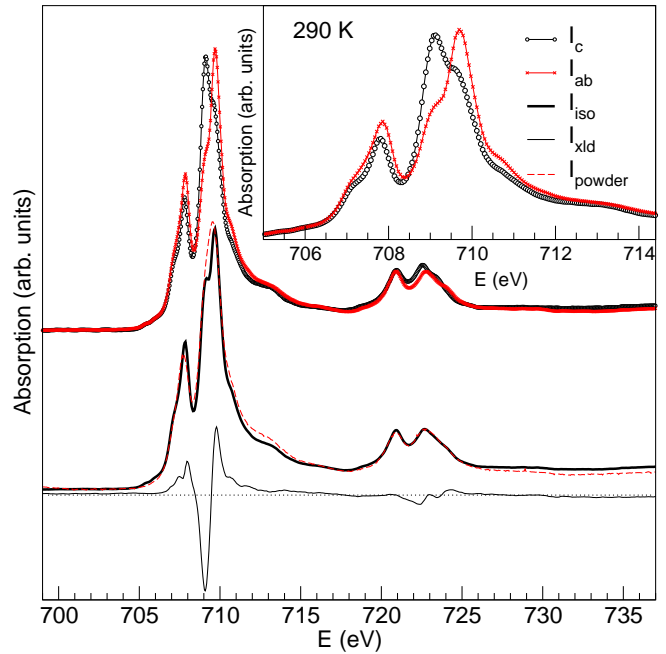}
\end{center}
\caption{(color online) x-ray absorption spectra of potassium iron
jarosite at 290 K measured with x-ray polarization parallel and
perpendicular to the $c$ axis, $I_{\text{c}}$ (black circles) and
$I_{\text{ab}}$ (red crosses), respectively. The lower trace shows the
corresponding values of the isotropic absorption, $I_{\text{iso}}=
\frac{2}{3}I_{\text{ab}}+\frac{1}{3}I_{\text{c}}$ (black line) and the
x-ray linear dichroism,  $I_{\text{xld}} = I_{\text{ab}}-I_{\text{c}}$
(thin black line). Also shown $I_{\text{iso}}$ from a powder sample
measured on 5U.1 at SRS Daresbury with a lower energy resolution
(broken, red line).  The inset shows a close up of $I_{\text{c}}$ and
$I_{\text{ab}}$ at the $L_3$ edge.}
\label{figure:xld1}
\end{figure}

Figure~\ref{figure:xld1} shows the spectra obtained
from single-crystal and powder samples of potassium iron jarosite at
290~K. The upper curves in Fig.~\ref{figure:xld1} display the x-ray
absorption spectra  from a single crystal measured on beamline ID08 at
the ESRF, with the incident x-ray  polarization parallel to the $c$
axis, $I_c$ (black circles), and perpendicular to the $c$ axis,
$I_{\text{ab}}$ (red crosses). The curves in the middle of the figure
give the corresponding isotropic absorption spectrum $I_{\text{iso} }=
\frac{2}{3}I_{\text{ab}} + \frac{1}{3}I_c$ (black line) and the
isotropic spectrum directly measured from a powder sample on station
5U.1 at the SRS (broken red line). The  curve at the bottom gives the
x-ray linear dichroism $I_{\text{xld}}$ defined as
$I_{\text{ab}}-I_{\text{c}}$ (thin black line). The $2p \to 3d$
transition consists of two edges, the first one at $\sim$$709$~eV
($L_3$ edge) corresponds to the creation of a $2p_{3/2}$ core hole.
The transition giving rise to a $2p_{1/2}$ core hole ($L_2$ edge) is
at $\sim$$12$~eV higher photon energy due to the $2p$ spin-orbit
coupling. The inset of Fig.~\ref{figure:xld1} shows the $I_{\text{c}}$
and $I_{\text{ab}}$ spectra in close up at the $L_3$ edge, which is
the most significant region of the spectrum.

\begin{figure}[htbp]
\begin{center} \epsfig{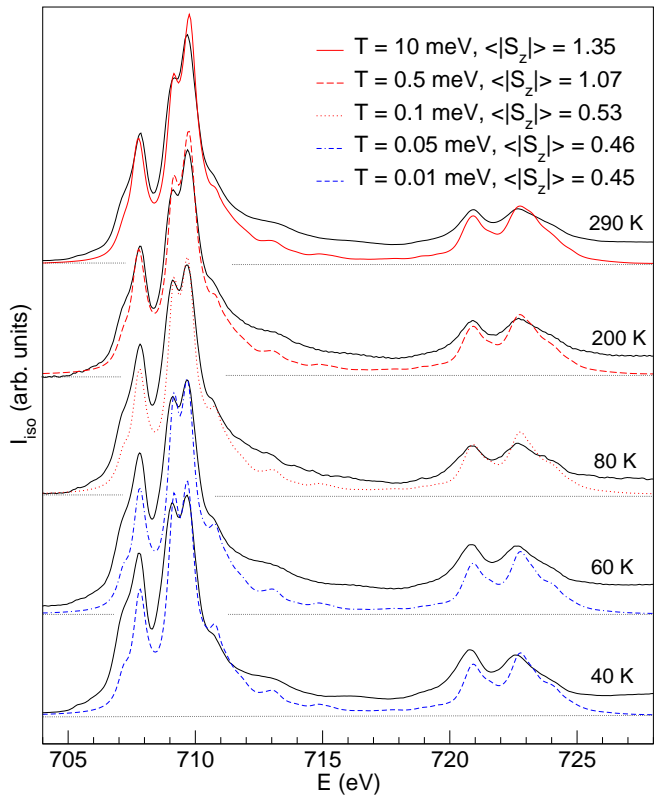}
\caption{\small{(color online) Temperature dependence of the isotropic
x-ray absorption measured at the Fe $L_{2,3}$ edges from a synthesized
potassium iron jarosite single crystal (black lines), compared to
calculated spectra (as discussed in Secs. IV and V) with simulated
temperatures as listed in the legend.}}
\label{figure:iso1}
\end{center}
\end{figure}

\begin{figure}[htbp]
\begin{center} \epsfig{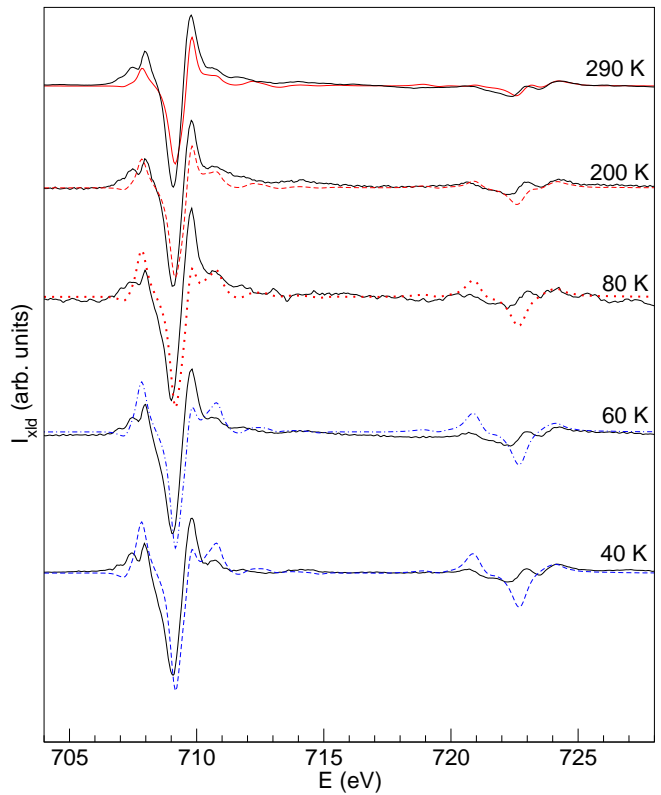}
  \caption{\small{(color online) Temperature dependence of the x-ray
linear dichroism measured at the Fe $L_{2,3}$ edges from a synthesized
potassium iron jarosite single crystal (black lines), compared to
calculated linear dichroism corresponding to the isotropic simulated
spectra of Fig.~\ref{figure:iso1}, scaled in intensity to fit the
experimental spectra by factors ranging from $1/2$ at 290~K down to
$1/3$ for the 60 and 40~K data. Possible reasons for this mismatch are
discussed in the text.}}
\label{figure:xld2}
\end{center}
\end{figure}

As seen in Fig.~\ref{figure:iso1}, the structure at the $L_3$ edge
becomes more prominent in the isotropic spectrum as the temperature is
lowered. At each stabilized temperature, the reproducibility of up to
eight spectra demonstrated the absence of any temporal variation in
the spectral line shape or intensity, therefore charging effects and
drift in sample alignment can be eliminated as the cause of the
observed temperature dependence. These isotropic spectra obtained by
weighted averaging of $I_{\text{ab}}$ and $I_{\text{c}}$ are in good
agreement with the spectra from powder samples measured for the SRS at
all temperatures. That this change in line shape occurs in the powder
averaged spectrum might be an indication of a slight change in the
electronic structure as the temperature is lowered. Only very small
changes were observed in the XLD spectral line shape on cooling the
sample, as shown in Fig.~\ref{figure:xld2}. This change in the
spectral line shape is accompanied by a slight reduction in the
integrated XLD spectrum~\cite{Scho:98}
\begin{equation} \int I_{\text{xld}}(\omega)d\omega /\int
I_{\text{iso}}(\omega)d\omega
\end{equation}  corresponding to a small change of the total
quadrupole moment of 0.05(2).  This might be interpreted as that the
Fe $3d$ shell gains between 3 and 7\% net $z^2-r^2$ character at base
temperature.

The room temperature spectra of hydronium iron jarosite were identical
to the spectra obtained on potassium iron jarosite within the
experimental resolution as measured at the SRS (the red, broken line
in the main panel of Fig.~\ref{figure:xld1} for comparison). 

\section{Ligand-field multiplet calculations} Due to the strong
electrostatic interactions between the $2p$ core hole and the $3d$
levels in the final state, the $2p$ $\to$ $3d$ absorption spectrum is
not simply proportional to the density of unoccupied $3d$ levels
\cite{vanderlaan:91} as a function of energy. For ferromagnets and
magnetically soft materials, where all the magnetic moments can be
aligned by applying an external field, the x-ray magnetic circular
dichroism (XMCD) and x-ray magnetic linear dichroism (XMLD) sum rules
can be applied to obtain the expectation values of the orbital and
spin magnetic moments \cite{Thole:92} and the magneto-crystalline
anisotropy \cite{vanderlaan:99}, respectively. In the case of the
jarosites it is not possible to align the magnetic moments parallel
with an external field due to the large energy scale of the
antiferromagnetic interactions ($\Theta_{\rm{CW}} \approx -800$
K). The only way to obtain accurate information about the relevant
expectation values is by explicitly calculating the transition
probabilities
\begin{equation} f(E_{\hbar\omega},\mathbf{q}) =
\frac{|\bra{\psi_g}\mathbf{\hat{r}_q}\ket{\psi_e}|^2}{E_{\hbar\omega}-\Delta
E - i\Gamma/2},
\end{equation} between the ground state $\psi_g$ and excited state
$\psi_e$, where $\mathbf{\hat{r}_q}$ is the electric-dipole operator
for x-ray absorption with polarization $\mathbf{q}$, and $\Gamma$ is
the lifetime broadening, which is treated as a fitting parameter. The
experimental x-ray spectra were used to obtain good approximations for
$\psi_g$ and $\psi_e$ with the aid of the \texttt {Hilbert++}
code~\cite{Mirone:00,Mirone:code}.  The Hilbert space for
$\ket{\psi_g}$ is spanned by determinants of the lowest atomic $LSJ$
configurations of $\ket{3d^n}$ and $\ket{3d^{n+1}\underline{L}}$,
where $\underline{L}$ denotes a ligand hole.  Likewise, the Hilbert
space for $\ket{\psi_e}$ is spanned by determinants of the atomic
configurations of $\ket{2p^5 3d^{n+1}}$ and $\ket{2p^5
3d^{n+2}\underline{L}}$.  The Slater integrals for the $3d$-$3d$ and
$2p$-$3d$ interactions and the $2p$ and $3d$ spin-orbit parameters for
the ground state and final state configurations calculated using
Cowan's atomic multiplet program~\cite{Cowan:81} are tabulated in
Ref.~\onlinecite{vanderlaan:92}. Approximate solutions for the
eigenfunctions of the iron ion in the compound  are then obtained by
diagonalization of the Hamiltonian including charge-transfer and
crystal-field terms arising from the surrounding oxygen
ligands~\cite{Mirone:code}
\begin{eqnarray} \mathcal{H} &=& \mathcal{H}_{\text{atom}}  \nonumber
\\  &+& \sum_{b}[t_{\sigma,b}d^+_{3\tilde{z}^2-r^2}p^-_{\tilde{z}}
+t_{\pi ,b}(d^+_{\tilde{x}\tilde{z}}p^-_{\tilde{x}} +
d^+_{\tilde{y}\tilde{z}}p^-_{\tilde{y}})+ \text{cc}] \nonumber \\
&+&\sum _{b} [V_{\sigma,b} d^+_{3 \tilde{z} ^{2}-r^2}d^-_{3
\tilde{z}^2-r^2}  +V_{\pi,b}  (d^+_{ \tilde{x} \tilde{z}} d^-_{
\tilde{x} \tilde{z}}+ d^+_{ \tilde{y} \tilde{z}}d^-_{ \tilde{y}
\tilde{z}})]  \nonumber \\  &+& \epsilon _p
\sum_{b}(p^+_{\tilde{x}}p^-_{\tilde{x}} +
p^+_{\tilde{y}}p^-_{\tilde{y}}  + p^+_{\tilde{z}}p^-_{\tilde{z}}). 
\end{eqnarray}  In this Hamiltonian, the first term represents the
atomic Hamiltonian. The second term gives the hybridization between
metal $3d$  and  oxygen $2p$ orbitals, where $t_{\sigma,b}$ and
$t_{\pi,b}$ are the Slater-Koster hopping parameters. For each bond,
$t_{\sigma,\pi}$ are rescaled with the bond length using
$t_{\sigma,\pi,b}=t_{\sigma,\pi}(R_{\text{b}}/R_{\text{ref}})^{\alpha}$
where $R_b/R_{\text{ref}}$ is the normalized length of Fe-O bond $b$
and the rescaling exponent $\alpha$ is treated as a free
parameter. The third term gives the contribution of the electrostatic
crystal field which was not used in our case. The last term gives the
energy of the $2p$ electrons in the oxygen valence band, where
$\epsilon _p$ is a simulation parameter related to the charge-transfer
gap $\Delta _{pd}$ via $\Delta _{pd} = U_{3d^5}  - \epsilon _{p}$. The
total $dd$ Coulomb energy for the addition of a sixth electron in the
Fe $3d$ shell $U_{3d^5}=\sum _{n=1}^5 U_n$ was found to be 44.3~eV by
considering the case $t_{\sigma,\pi}=0$, where no charge transfer
occurs unless the Fe$^{2+}$ $d^6\underline{L}$ configuration has a
lower energy than the Fe$^{3+}$ $d^5$ configuration. The sum is made
over all bonds $b$, where the local coordinate frame
$\tilde{x},\tilde{y},\tilde{z}$ is oriented with the $\tilde{z}$ axis
along the bond direction. Numerous comparisons between experiment and
theory~\cite{Thole:PRB85,FdG:05Rev} have shown that this approach
works well, provided that the Slater integrals for the Coulomb
interactions and spin-orbit couplings are reduced to 70-80\% of the
atomic Hartree-Fock values \cite{Thole:88Spinmix}. Here, all $pd$ and
$dd$ Slater integrals were reduced to 80\% of the atomic values
calculated using Cowan's program. The lifetime broadening $\Gamma$ was
set to $0.24$~eV for the $L_3$ edge and $0.3$~eV for the $L_2$
edge. These are values as commonly used in the calculation of x-ray
absorption spectra, with $L_2$ always slightly larger than $L_3$. This
spectral broadening is due to the coupling with the (infinite
dimensional) electromagnetic field. 

\begin{figure}[htbp]
\begin{center} \epsfig{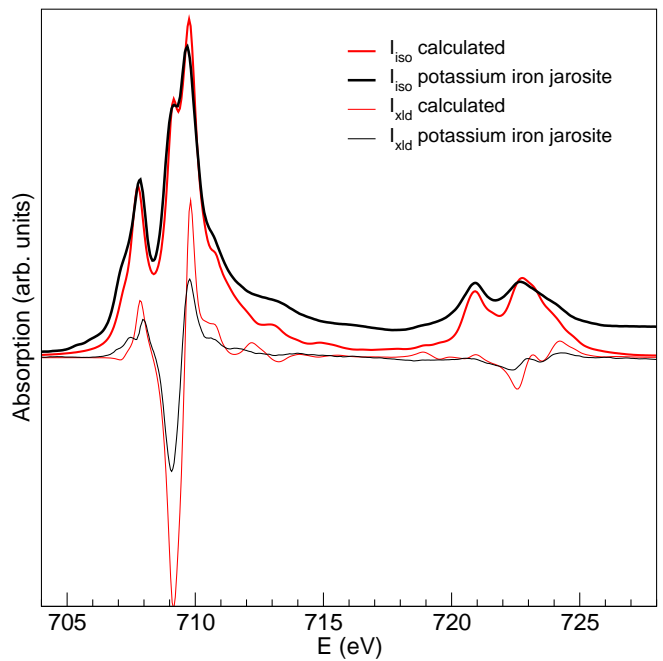}
\caption{(color online) Experimentally obtained isotropic spectrum and
x-ray linear dichroism (black lines) at 290 K and simulated XAS and
XLD (red lines) Boltzmann averaged at 110 K. The simulated XLD is
roughly twice as large as the experimentally obtained XLD which is
ascribed to mis-alignments and sample imperfections.}
\label{figure:LFcalc}
\end{center}
\end{figure}
  
The crystal field at the iron atom in potassium iron jarosite is
approximately of  $D_{4h}$ symmetry, with an elongation of the bond
length  $R_{\text{ap}}$ along the fourfold symmetry axis ($c'$) of
$\sim$4\% with respect to the bond length  $R_{\text{eq}}$ in the
equatorial plane~\cite{Grohol:03}. The local symmetry axes of the
FeO$_6$ octahedra are canted with respect to the crystallographic $c$
axis by an angle of 19.2$^{\circ}$, resulting in buckled FeO$_4$
planes forming the kagome layers. 

Our simulation shows that the spectra are mainly determined by the
strongly anisotropic $\sigma$-type hybridization between the iron $3d$
and oxygen $2p$ orbitals. The electrostatic part of the crystal field
could be left out. An excellent agreement between experimental  and
simulated spectra for $I_{iso}$ was obtained using only four
simulation parameters, namely the charge-transfer gap $\Delta _{pd}$,
$t_{\sigma}$, $t_{\pi}$ and the rescaling constant $\alpha$ as defined
in $t_{\pi,\sigma}^{\text{ap}} = t_{\pi,\sigma}^{\text{eq}}
(R_{\text{ap}}/R_{\text{eq}})^{\alpha}$ (with
$R_{\text{ap}}/R_{\text{eq}} = 1.035$ in the present case).  When the
simulated isotropic spectra is in good agreement with the experimental
spectrum, at the same time optimal agreement is obtained for the
linear dichroism spectrum. At high temperature there is good agreement
between the experimental and simulated linear dichroism spectrum,
apart from an intensity scaling factor of 0.5 as shown in
Figs.~\ref{figure:xld2} and \ref{figure:LFcalc}. With the fitting
parameters for the simulation there is no way to reduce the overall
intensity of the linear dichroism without radically modifying the
isotropic spectrum. For up to 6\% the intensity difference might be
attributable to the local canting of the FeO$_6$ octahedra with
respect to the $ab$-plane. In addition to this there is an inevitable
loss of intensity in the XLD due to misalignments and sample
imperfection. Reduced energy resolution can also lead to a reduction
of the measured XLD.  We conclude that the difference in overall
intensity of the experimental and calculated $I_{\mathrm{xld}}$ is not
due to fundamental shortcomings of the simulation.  For reasons we do
not quite understand, the agreement between theory and experiment as
evident in the temperature dependence of the isotropic spectra is not
reflected in the XLD spectra at lower temperatures. We expect that
these spectra, which are averages over increasingly fewer states, are
also increasingly sensitive to shortcomings in our simulations. Since
the XLD is an orientational energy dependence it will be particularly
sensitive for any shortcomings in our theoretical model. It should be
emphasised that, despite the differences between the experimental and
simulated spectra at low temperatures, the overall agreement between
experiment and simulation as obtained here is remarkably good. The
parameters as found in our simulation are given in
Table~\ref{table:resxld} (left-hand side). The large error bar on the
value for the charge-transfer gap of 3.5(2.0) eV reflects the
adjustment of the total $dd$ Coulomb energy on the iron site
$U_{dd}^{\text{tot}} = U_{3d^5}(\langle n \rangle -5)$ to any increase
in $\epsilon _p$ by transfer of additional electron weight into the Fe
ion. This is reflected in the total error in $\langle n \rangle$ of
$\sim$0.1 electron.  

The values
found here for $\Delta_{pd}$ and $t_{\sigma}$ are similar to those
found previously for other Fe$^{3+}$ compounds~\cite{FdG:03JACS, FdG:05Rev}.  
Fe$_2$O$_3$, which is in many respects probably the compound most
closely related to iron jarosite, was previously found to have a band
gap of predominantly charge-transfer character of 2 to 2.5
eV~\cite{Ciccacci:91}. 
With an unusually small value for $t_{\pi}$ of 0.3(2) eV the lowest
energy peak, at 707.8 eV is reproduced in the simulated spectrum. The
main peak at 709 eV splits into two peaks with the introduction of a trigonal
distortion of the FeO$_6$ octahedron. A common value for the rescaling
exponent $\alpha$ is $-$3 and the small value found here ($-$7.5) is an
indication that the charge anisotropy is larger than the crystal field
anisotropy as defined by the oxygen positions around the iron. The
lowest energy low-spin configuration, corresponding to the
$^4F_{11/2}$ free-ion configuration, lies 1.2 eV higher in energy and
is not found to mix into the ground state. 

\begin{table}[ht]
\caption{\small{Results from simulations of the Fe $L_{2,3}$
  spectra of potassium iron jarosite, with ligand-field multiplet
  calculations using the  \texttt{Hilbert++}
  code\cite{Mirone:code}. The left-hand side  lists all the parameters
  varied in the simulation. 
  The right-hand side gives the resulting expectation values for
  the calculated ground-state properties of the Fe ion. The errors indicate
  the range over which the parameters could be varied whilst maintaining
  a good agreement with the data. The right-hand side of the lowest panel
  gives the effective  moment in the case of full orbital
  polarization, while the calculation indicates only a very weak
  orbital polarization of the added electron weight due to charge
  transfer. We use the convention that upper case $J,L,S$ represent operators
  and lower case $j,l,s$ represent the quantum numbers given by
  $\langle L^2 \rangle = l(l+1)$ etc. 
 }}
 \label{table:resxld}

\begin{tabular}{lr|lr}
\hline\hline
Simulation pars. && Expectation values&\\
\hline
$\Delta _{pd}$ (eV) & 3.5(2.0) & $D_z$ (meV) & 0.5(1)\\
$t_{\pi}$ (eV) & 0.3(2)& $\sqrt{\langle S^2 \rangle}=\langle S\rangle$ & 2.77(6) \\
$t_{\sigma}$ (eV) & 3.0(2)& $\leadsto s = 2.32$&\\
$\alpha$& $-$7.5 & $\sqrt{\langle L^2 \rangle}(\neq \langle L\rangle)$ &
1.5(2)\\
& & $\leadsto l = 1.11$&\\
& & but $\langle L \rangle$ & 0.008(3)\\
& & $\langle n \rangle$ & 5.39(7)\\
\hline
From $t_{\sigma,\pi}$ and $\alpha$:& (in eV)&Using $\langle L \rangle$ we obtain&\\
$t_{\sigma}^{\text{eq}}$ &3.0(2)&$j = s  (+l\approx 0)$& 2.3\\
$t_{\sigma}^{\text{ap}}$ & 2.3(2)&$g_{J}$ & $\sim 2$\\
$t_{\pi}^{\text{eq}}$ &$<0.5$&$\mu_{\text{eff}}$ & $5.5(1)\mu_{\text{B}}$\\
$t_{\pi}^{\text{ap}}$ &$<0.4$&and using $\sqrt{\langle L^2 \rangle}$\\
&&$j = s + l$& 3.4\\
&&$g_{J}$ & 1.67(5)\\
&&$\mu_{\text{eff}}$ & $6.51(7)\mu_{\text{B}}$\\
\hline \hline

\end{tabular}

\end{table}

The single-ion anisotropy $D_z$ is obtained as the difference between
the energy of the $S_z$ =  $\pm 1/2$ and $\pm 5/2$ levels and was
found to be of the easy-plane type. This easy-plane anisotropy was
also confirmed by rotating the spin quantisation axis to various
directions with respect to the crystal field. In this way we have also
attempted to obtain estimates for $E_{xy}$. The $D_{4h}$ symmetry of
the ligand field was lowered to $C_2$, to model the actual positions
of the oxygen anions around  the cation as obtained from
crystallographical data~\cite{Grohol:03,WillsKjar:00}. However, this
had no effect on the agreement of the calculated spectra with
experiment, nor on the values as listed in Table~\ref{table:resxld}
and hence we consider an estimate for $E_{xy}$ from our data presently
not attainable. It should in this respect be noted that the value for
$D_z$ was found to change with the splitting of the main peak at
709~meV, introduced by the trigonal distortion of the ligand
field. Its value is therefore tightly constrained by the
experimentally obtained spectra and the 3d spin-orbit coupling
strength.  As is detailed in Table~\ref{table:resxld} our calculation
indicates that for the isolated FeO$_6$ cluster most of the orbital
angular momentum of the Fe 3$d^6$ configuration is quenched by the
ligand field, with $\langle L \rangle = 0.008(3)$ while the maximum
orbital angular momentum which could in this case be realised as
$\sqrt{\langle L^2 \rangle} = 1.5(2)$. 

Due to the lower resolution at which the hydronium jarosite powder
spectrum has been measured we cannot provide detailed information
about this system, apart from a confirmation that the general crystal
field is similar, with comparable charge-transfer and ($\sigma$-type)
hybridization. This is a good indication that the magnetic exchange in
hydronium jarosite is likely to be comparable with that of the other
jarosites, i.e. with a Weiss temperature around $-$800~K.  There is also
a slight evidence that the $D_{4h}$ ligand-field anisotropy and hence
the zero-field splitting, is slightly smaller in hydronium jarosite.

\section{Discussion}
The expectation values of the ground-state obtained in our calculation
are given in Table~\ref{table:resxld} (right-hand side). The
corresponding free-ion configuration is $d^5$ $^6S_{5/2}$ with an
additional electron weight of $\sim$0.4 electron that is mainly in the
$d(x^2-y^2)$ orbital. The spin-orbit coupling is much smaller than the
$t_{\sigma}$ hopping bandwidth and hence the fraction of the electron
transferred to the ligands to the Fe cation is only very weakly
orbitally polarized, with $\langle L \rangle = 0.008(3)$. This value
is much smaller than the orbital angular momentum as inferred from
high-temperature magnetic susceptibility data, which indicate that
$\mu_{\text{eff}} \approx 6.5$ $\mu_{\text{B}}$~\cite{Grohol:05}. As shown in
the lower panel of Table~\ref{table:resxld} (right-hand side) the
experimental value is remarkably close to the case where the added
electron weight due to charge transfer is fully orbitally
polarized. It is not expected that in this $3d$ transition metal
compound $l$ and $j(=s+l)$ are good quantum numbers, but there is sufficient
evidence that $\langle L \rangle$ is significantly larger than
zero~\footnote{Iron jarosite is in this sense not a unique case. For
  $3d$ transition metal compounds with more than half-filled shells
  (such as e.g. Herbertsmithite~\cite{deVries:07}) the effective
  moments are in general significantly larger than the spin-only
  moment.}. 
It might be that comparison of experimental results with the Mott-Hubbard-like
calculations as described here, but with larger clusters containing
two or three iron centres, can provide an explanation for the
discrepancy between the calculated and experimentally obtained orbital
angular momentum.  

\begin{figure}[htbp]
\begin{center}
  \epsfig{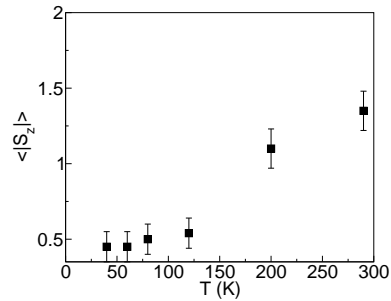}
\caption{The expectation values $\langle |S_z| \rangle$ of the
  simulated spectra, against the temperature of the corresponding
  experimental spectrum. For a single $S=5/2$ spin the minimum
  for $\langle|S_z|\rangle $ is 0.5 and the maximum is 1.5 (equal
  population of all $m_S$. The gradual increase of
  $\langle|S_z|\rangle $ with temperature is in agreement with earlier
  neutron measurements which show a gradual alignment of the spins in
  the kagome plane at temperatures well above the transition
  temperature of 64~K~\cite{Grohol:05}. }
\label{figure:Sz}
\end{center}
\end{figure}

The energy of $D_z$ = 0.5(1)~meV for the easy-plane single-ion
anisotropy is in good agreement with the values found by
Matan~\emph{et al.}~\cite{Matan:06} (0.428(5)~meV) and Coomer~\emph{et
al.}~\cite{Coomer:06} (0.47(2)~meV) in their spin-wave analysis of
inelastic neutron data from single crystals and powders,
respectively. This implies that an easy-plane anisotropy, along with
further neighbor interactions, can explain the magnetic ground state
in potassium iron jarosite. Further evidence might be found in the
temperature dependence of the experimental and calculated isotropic
spectra.  The spectra in Fig.~\ref{figure:LFcalc} were taken at room
temperature and were compared with the calculated spectra for the
Boltzmann averaged configurations of the FeO$_6$ cluster.  As is
visible in Fig.~\ref{figure:iso1} the relative intensity of the two
maxima in the split peak at 709 eV changes in a similar way as was
observed in the experimentally obtained spectra. At the same time an
(unexplained) increase in intensity of the peak at 707.8 eV peak
relative to the intensity of the main peak is observed in the
experimental spectra. We have not been able to account for the latter
(which was also observed in powder samples) either by changing the
temperature of the calculated spectra or with simulations with new
values for $\delta _{pd}$, $t_{\sigma,\pi}$ and $\alpha$. The change
in the main peak at 709~eV where we obtain good agreement, is a
direct result of the spin-orbit coupling which causes a change in the
shape of the electronic $3d$ orbital as the spins align in the easy
plane. The change in shape of the main peak in the experimental
spectra was found to occur between 200~K and 40~K. This is in
agreement with previous experiments which have shown that in potassium
jarosite the alignment of spin with the kagome planes sets in above
120~K, i.e. well above the transition temperature of 65
K~\cite{Grohol:05}.  The temperature dependence in the experimentally
obtained spectra was simulated with the temperature (given in units of
eV) used for the Boltzmann averaging over excited states as the only
extra parameter, as shown in Fig.~\ref{figure:iso1}. All other
simulation parameters were fixed to the values given in
Table~\ref{table:resxld}. In this way we found the expectation values
for $\langle |S_z|\rangle$ at each corresponding experimental
temperature, arising from the thermal population of the $m_S$ levels
in the $S=5/2$ multiplet. The temperature dependence of
$\langle |S_z|\rangle$ (Fig.~\ref{figure:Sz}) shows a gradual
alignment of the spins into the kagome planes, at temperatures well
above the transition temperature to a long-range ordered state at
64~K. This is in rough agreement with the degree of co-planarity of the
spins as measured using neutron diffraction on large single
crystals~\cite{Grohol:05}. 

The simulated linear dichroism spectra at the
lowest temperatures, corresponding to states with the spins aligned in
the equatorial plane (the states $\ket{s,m_{z'}}= \ket{2.5,\pm 0.5}$)
do not agree well the experimentally obtained spectra as shown in
Fig.~\ref{figure:xld2}, although some features in the temperature
dependence of the XLD are reproduced in the simulated spectra. Without
success we have attempted to improve on this situation by calculating
the spectra for the Fe spin aligned along the $x'$ axis in the
equatorial plane (i.e. $\ket{s,m_{x'}}= \ket{2.5,\pm 2.5}$) which
resembles more closely the actual ground state in potassium
jarosite. The remaining disagreement could be in part due to
experimental error due to sample imperfections and in part due to
intrinsic limitations of the simulation. 

It might be naive to expect that the ligand-field can be inferred from
crystallographic data alone. A first indication for this is the large
value of $\alpha$ obtained here, which amplifies the effect of the
crystallographic trigonal distortion of the FeO$_6$ octahedra. Even
when the symmetry of the ligand field is further lowered from
$D_{4h}$ and accurately modelled on the locations of the oxygen
anions around the central iron~\footnote{Though all Fe-O$_{\text{eq}}$
  bond lengths are the same, the equatorial plane is rectangular rather
  than square, with the short side pointing into the kagome
  triangles. Furthermore, the Fe-O$_{\text{ap}}$ axis is canted by
  $4^{\circ}$ away from the local $c'$ axis.} neither the calculated
spectra nor the simulation parameters change significantly.  Therefore,
it is possible that the actual charge anisotropy follows more closely
the $xy$ plane rather than the equatorial oxygens. The easy-plane
magnetic anisotropy of $D_z = 0.5(1)$~meV found here (see
Table~\ref{table:resxld}), is then not directly related to the
trigonal distortion of the crystal field but more to the 2D
  character of the electronic hopping and magnetic exchange, the
virtual absence of $\pi$ symmetry hybridization and the Fe $3d$
spin-orbit coupling strength. In that case the easy plane could
coincide more closely with the kagome plane. The strong 2D character
can in part be attributed to the sulphate groups separating the
kagome layers. The formal valence of the sulphur cation is 6+
with the [Ne] electronic configuration, but this large electrostatic
charge is likely to attract electron weight from the neighboring
oxygen anions, including the oxygens apical in the FeO$_6$ octahedra. 
The hopping parameters $t_{\sigma,\pi}$ from these oxygens to the iron is
then further reduced with respect to the in-plane hopping. This
translates into the large value for $\alpha$ found here.

The large difference between $t_{\pi}$ and $t_{\sigma}$ obtained here
might be explained by the triangular arrangement of the neighboring
FeO$_6$ octahedra. This dramatically lowers the symmetry of the 
local crystal field around the oxygen ligands and reduces the $\pi$
symmetry overlap of the oxygen $2p$ orbitals with the iron $3d$
orbitals. Due to the low symmetry of the oxygen sites the symmetry of
the crystal field can be expected to be even lower than indicated by
the crystallographic positions of the anions. It is in this sense
fortunate that the situation can be modelled so well by the
anisotropic $\pi$ and $\sigma$ hopping whose ratio is equal for all six
anions around the central iron. It could well be that the last differences
between the calculated and experimental spectra, such as the low
temperature x-ray linear dichroism, some of the difference in the
overall intensity of the XLD and the missing bump just above 707 eV,
could be brought into agreement by allowing a different balance
between $t_{\pi}$ and $t_{\sigma}$ for the apical oxygen anions.

In neutron spin-wave studies~\cite{Matan:06, Yildirim:06,Coomer:06}
the crystal field anisotropy scenario has been compared in close
detail with that of DMI. In the present context such a discussion
would require calculations on clusters containing two to three iron
centres. Such a calculation would certainly be of great interest, but
has not been attempted here, since the present approach already
accounts so well for the experimental observations on potassium iron
jarosite. The present result closes the last missing causal link, from
electronic structure to collective magnetic ground state.

\section{Conclusion}
The isotropic and x-ray linear dichroism spectra taken at the Fe
$L_{2,3}$ edges in potassium iron jarosite indicate that an
orbital angular momentum and zero-field splitting arise due to an
anisotropic, mainly $\sigma$ type, charge-transfer from the oxygen
ligands. This closely resembles the scenario described
in Ref.~\onlinecite{Wan-Lun:94} to explain the zero-field splitting as observed
in Mn$^{2+}$ salts.

The zero-field splitting parameter of the crystal field anisotropy of
0.5~meV as found here is in excellent agreement with the value found from
fits of a crystal field anisotropy model to the spin-wave spectrum in
the ground state of potassium iron jarosite~\cite{Matan:06,
  Yildirim:06, Coomer:06}. However, the most important result obtained
here is that we have verified in detail the mechanism by which this
magnetic anisotropy arises, closing the remaining missing link between the
electronic structure in potassium iron jarosite and its magnetic
ground state. We have even measured a change in the electronic
structure as the magnetic ground state becomes co-planar, with the
change of the isotropic line shape over the relevant temperature
range. Of course, an additional in-plane anisotropy $E_{xy}$ and further-
neighbor interactions are needed to fully explain the magnetic ground
state. The present study does not provide more hints to the origin
of the glassy state in hydronium jarosite. We found that the energy
scale of the magnetic interaction in hydronium jarosite should be
comparable to that in potassium jarosite. It might be that temperature
dependent spectra taken at a higher resolution could provide new
insights in the case of hydronium jarosite. A final
interesting observation is that the maximum orbital angular momentum
polarization possible for the amount of charge-transfer observed here,
but not the actually calculated orbital polarization, is in rough
agreement with the experimentally observed effective moment. 

\section{Acknowledgement}
Simon Parsons and Clivia Hejny (Edinburgh University) are acknowledged
for their help with the characterization and alignment of single
crystals of potassium jarosite. We would like to thank Nicholas
Harrison (Imperial College, London and Daresbury Laboratory) and Barry
Searle (Daresbury Laboratory) for fruitful discussions about LSDFT
calculations on potassium jarosite. Peter Bencok (ESRF and Diamond) and Nicola
Farley (Daresbury Laboratory)
are acknowledged for their help during x-ray spectroscopy
measurements. We further thank Claudine Lacroix (CNRS, Grenoble), Andrew
Wills (UCL), Frank de Groot (Utrecht University) and Paul Attfield
(Edinburgh University) for fruitful discussions. MdV gratefully
acknowledges financial support from the Centre for Materials Physics and
Chemistry (Science and Technology Facility Council, UK) and the Highly Frustrated Magnetism network of the
European Science Foundation.

\end{document}